# Non-Relativistic Quantum Particle on the Surface of the Cylinder under the Stark-like Perturbative Potential

Deriyan Senjaya

*Department of Physics, National Tsing Hua University, Hsinchu 30013, Taiwan*

Email: d_senjaya@gapp.nthu.edu.tw

**Abstract**

The Kaluza-Klein (KK) theory is a well-known theory that brings an idea to unify the electromagnetic interaction with the gravitational interaction by introducing the concept of extra dimension [1, 2, 3]. In the KK theory, the extra dimension is somewhat cloaked and capable to describe why gravitational interaction is weaker than the other interactions. In addition, uncloaking this extra dimension needs high energy to observed. It is interesting if one can learn how to uncloak this extra dimension and learn their existence in an easy way. By this motivation, this research investigates the effect of Stark-like perturbative potential $\hat{H}_{SL} = \beta z V_{o_z}(\theta)$ to the non-relativistic quantum particle on the surface of the cylinder (QPSC) (with length of $L$ and radius of $R_o$) [4, 5] using the perturbation theory. The QPSC is chosen due to its similarity to the framework of KK theory (where the extra dimension is encoded to the angular variable $\theta$). The Stark-like perturbative potential is chosen from the motivation of hydrogen atom energy level splitting. The energy level splitting by this potential on the QPSC will be interpreted as the uncloaked extra dimension. The result of this research shows that Stark-like perturbative potential is effective to split the QPSC energy level in the degenerate case ($R_o = L/\pi$). This becomes effective because in the degenerate case, the first-order energy correction depends on the QPSC quantum numbers $n_z$ and $n_\theta$.

**Keywords**: Kaluza-Klein Theory, Extra Dimension, Quantum Particle on the Surface of the Cylinder, Stark-like Perturbative Potential, Perturbation Theory

## INTRODUCTION

The combination of gravitational and electromagnetic interactions is well-described by the Kaluza-Klein (KK) theory by introducing the concept of extra dimension ($x^5$). The $x^5$ in KK theory is somewhat cloaked and represents the argument why gravitation is weaker than electromagnetic [1, 2, 3]. In addition, to observe the existence of the $x^5$ in KK theory, one should give a high-energy scale perturbation [3]. The need for high-energy scale perturbation will affect the cost of the actual experiment, e.g., in the Large Hadron Collider (LHC). This fact obstructs many physicists from answering the question of whether the extra dimension exists or not.

The process of how to uncloak the $x^5$ in KK theory into reality becomes one of the meaningful findings in the scope of fundamental physics. Therefore, this paper tried to consider the behavior of the Quantum Particle on the Surface of a Cylinder (QPSC) under Stark-like perturbative potential as a solution to understand how to uncloak the extra dimension in a simple way using a simple quantum system. The QPSC is a simple quantum system that mimics the behavior of the KK theory [4, 5]. Meanwhile, the perturbative potential selection as the Stark-like potential is motivated by the idea of energy level splitting in the hydrogen atom [6, 7]. The effect of Stark perturbative potential in the hydrogen atom splits the energy level degeneracy. When the Stark-like perturbative potential is on to the QPSC, one may have the splitting effect like the hydrogen atom and interpret it as the uncloak process to see the extra dimension.

# A QUANTUM PARTICLE ON THE SURFACE OF A CYLINDER (QPSC)

Consider a non-relativistic free particle of mass *m* confined on the surface of a cylinder with radius $R_o$ and length *L*, as shown in Fig. 1. In the quantum mechanics picture, the behavior of the particle in this system is well-described by the Schrodinger equation in Eq. (1) using the cylindrical coordinates $(r, \theta, z)$ with $\hat{V} = 0$ ($0 \leq z \leq L$ and $0 \leq \theta \leq 2\pi$).

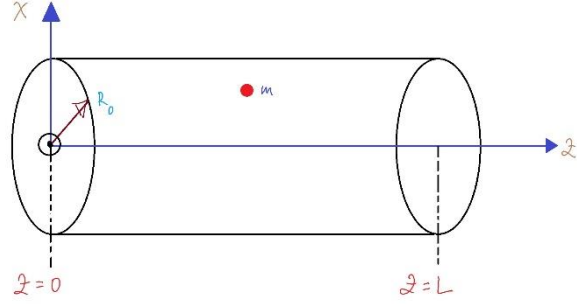

**FIGURE 1.** Quantum particles on the surface of a cylinder

$$-\frac{\hbar^2}{2m}\left[\frac{1}{r}\frac{\partial}{\partial r}\left(r\frac{\partial \Psi}{\partial r}\right) + \frac{1}{r^2}\frac{\partial^2 \Psi}{\partial \theta^2} + \frac{\partial^2 \Psi}{\partial z^2}\right] = E\Psi \quad (1)$$

Here, the $\Psi = \Psi(r, \theta, z, 0)$ indicates the particle's wave function at $t = 0$, and $E$ is the particle's energy. To solve Eq. (1), one can use the trial function (*ansatz*) of variable separation in the form of the following:

$$\Psi(r, \theta, z, 0) = R(r)\Theta(\theta)Z(z) \quad (2)$$

The $R(r)$ is a function that only depends on $r$, $\Theta(\theta)$ is a function that only depends on the $\theta$, and $Z(z)$ is a function that only depends on *z*. Substituting Eq. (2) into the Schrodinger equation and dividing all sides by $R\Theta Z$ and $-\hbar^2/2m$, one can obtain:

$$\frac{1}{rR}\frac{d}{dr}\left(r\frac{dR}{dr}\right) + \frac{1}{r^2\Theta}\frac{d^2\Theta}{d\theta^2} + \frac{1}{Z}\frac{d^2Z}{dz^2} = -\frac{2mE}{\hbar^2} \quad (3)$$

For the free particle on the surface of a cylinder, there is a fact that helps reduce the complexity of Eq. (3). The particle on the surface of a cylinder always has a constraint of motion at $r = R_o$. Therefore, the $\Psi$ only depends on two variables, $\theta$ and $Z$. Let $R(r)$ be a constant, then the differential form of $dR/dr$ in Eq. (3) will vanish, and Eq. (3) turns out to be Eq.(4) as the following:

$$\frac{1}{\Theta}\frac{d^2\Theta}{d\theta^2} + \frac{R_o^2}{Z}\frac{d^2Z}{dz^2} = -\frac{2mER_o^2}{\hbar^2} \quad (4)$$

Let the first term of the left-hand and right-hand sides of Eq. (4) be a constant $U^2$ and $V^2$. Then, one can find two separable equations, as shown in Eq. (5) and Eq. (6) below.

$$\frac{1}{\Theta}\frac{d^2\Theta}{d\theta^2} = -U^2 \rightarrow \frac{d^2\Theta}{d\theta^2} + U^2\Theta = 0 \quad (5)$$

$$\frac{R_o^2}{Z}\frac{d^2Z}{dz^2} = -V^2 \rightarrow \frac{d^2Z}{dz^2} + \frac{V^2 Z}{R_o^2} = 0 \quad (6)$$

The Eq. (5) and Eq. (6) are two second-order ordinary differential equations (ODE) with the oscillation-like model. Therefore, the general solutions for Eq. (5) and Eq. (6) are:

$$\Theta(\theta) = C_1 e^{iU\theta} + C_2 e^{-iU\theta} \tag{7}$$

$$Z(z) = C_3 \sin\left(\frac{Vz}{R_o}\right) + C_4 \cos\left(\frac{Vz}{R_o}\right) \tag{8}$$

Based on Eq. (5) and Eq. (6) general solutions, one can conclude the general solution of $\Psi(\theta, z, 0)$ is expressed in Eq. (9) below. The complete wave function of the system can obtained using the appropriate boundary conditions.

$$\Psi(\theta, z, 0) = \left[C_3 \sin\left(\frac{Vz}{R_o}\right) + C_4 \cos\left(\frac{Vz}{R_o}\right)\right]\left[C_1 e^{iU\theta} + C_2 e^{-iU\theta}\right] \tag{9}$$

*First Boundary Condition*

The particle in the $\theta$-direction moves in clockwise (CW) or counterclockwise (CCW) direction only. The particle in this direction cannot move back and forth. Therefore, mathematically speaking, one can safely assume that $C_2 = 0$. By this boundary condition, Eq. (9) is reduced into Eq. (10) as follows:

$$\Psi(\theta, z, 0) = C_1 \left[C_3 \sin\left(\frac{Vz}{R_o}\right) + C_4 \cos\left(\frac{Vz}{R_o}\right)\right] e^{iU\theta} \tag{10}$$

The particle in the $\theta$-direction also moves periodically. So, one can find that the wave function Eq. (10) should fulfill the periodic boundary condition $\Psi(0, z, 0) = \Psi(2\pi, z, 0) \rightarrow e^{2\pi iU} = 1$. Using the periodic boundary condition, the solution of $U$ should be $n_\theta$. The $n_\theta$ is called the $\theta$-Quantum Number and belongs to the integer values ($n_\theta = 1, 2, 3, ...$). Therefore, one can rewrite Eq. (10) as the following.

$$\Psi(\theta, z, 0) = C_1 \left[C_3 \sin\left(\frac{Vz}{R_o}\right) + C_4 \cos\left(\frac{Vz}{R_o}\right)\right] e^{in_\theta \theta} \tag{11}$$

*Second Boundary Condition*

In the z-direction, the particle is in the $0 \leq z \leq L$. The particle is not allowed at $z = 0$ and $z = L$. Then, by this boundary condition, one can say that $\Psi(\theta, 0, 0) = \Psi(\theta, L, 0)$. As a result, one can also reduce and define the value of $VL/R_o = n_z \pi$. The $n_z$ is the z-Quantum Number and belongs to the integer ($n_z = 1, 2, 3, ...$).

$$\Psi(\theta, z, 0) = C_1 C_3 \sin\left(\frac{n_z \pi z}{L}\right) e^{in_\theta \theta} \tag{12}$$

*Calculating the coefficient $C_1 C_3$*

The wave function $\Psi(\theta, z, 0)$ should be normalized. The normalized wave function guarantees the existence of the particle in the system. Using the procedure for the normalization, one can find that:

$$\int_0^{2\pi} \int_0^L |\Psi(\theta, z, 0)|^2 (R_o d\theta) dz = 1 \rightarrow 2\pi R_o (C_1 C_3)^2 \int_0^L \sin^2\left(\frac{n_z \pi z}{L}\right) dz = 1 \tag{13}$$

Using $\cos(2x) = 1 - 2\sin^2 x$, one can find this following result:

$$2\pi R_o (C_1 C_3)^2 \left[\frac{L}{2}\right] = 1 \rightarrow C_1 C_3 = \frac{1}{\sqrt{\pi R_o L}} \tag{14}$$

Plugging back the $C_1 C_3$ coefficient in Eq. (14) into Eq. (12), the complete solution of the wave function is in the following relation (Eq. (15)).

$$\Psi(\theta, z, 0) = \frac{1}{\sqrt{\pi R_o L}} \sin\left(\frac{n_z \pi z}{L}\right) e^{i n_\theta \theta} \tag{15}$$

From the Eq. (15), one may construct the probability density of the particle as the following:

$$\rho(\theta, z, 0) = |\Psi(\theta, z, 0)|^2 = \frac{1}{\pi R_o L} \sin^2\left(\frac{n_z \pi z}{L}\right) \tag{16}$$

The probability density of the particle in Eq. (16) is independent of the $\theta$-direction. Eq. (16) is also similar to the probability density of the one-dimensional infinite potential well [8] with different constant. However, one already knows that the QPSC is a two-dimensional system. By Eq. (16), one can say that this $\theta$-dimension is cloaked, analogous to the $x^5$ in KK theory. Therefore, this is the first indication that the QPSC is the analogous model of the KK theory.

*Energy Solution*

By the solutions of $U$ and $V$, one can write the energy solution $E$. The energy $E$ depends on two quantum numbers and two geometrical parameters of the system ($L$ and $R_o$). The dependence of $E$ on those two quantum numbers indicates the system is quantized.

$$U^2 + V^2 = \frac{2mER_o^2}{\hbar^2} \tag{17}$$

$$E = E_{n_z, n_\theta}(L, R_o) = \frac{n_z^2 \pi^2 \hbar^2}{2mL^2} + \frac{\hbar^2}{2m}\left[\frac{n_\theta}{R_o}\right]^2 \tag{18}$$

The energy in Eq. (18) indicates the non-degenerate type of energy in general (for different $L$ and different $R_o$). That means for two distinct quantum states, one can find one value of the particle's energy (one-to-one correspondence) [8]. However, if one sets the specific condition for $R_o$ and $L$, such as $R_o = L/\pi$ and $L = \pi R_o$, the indication of the degenerate type of energy appears (see Eq. (19) and Eq. (20)).

$$E = E_{n_z, n_\theta}(L, L/\pi) = \frac{n_z^2 \pi^2 \hbar^2}{2mL^2} + \frac{\hbar^2}{2m}\left[\frac{n_\theta \pi}{L}\right]^2 = \frac{\pi^2 \hbar^2}{2mL^2}(n_z^2 + n_\theta^2) \tag{19}$$

$$E = E_{n_z, n_\theta}(\pi/R_o, R_o) = \frac{n_z^2 \pi^2 \hbar^2}{2m(\pi R_o)^2} + \frac{\hbar^2}{2m}\left[\frac{n_\theta}{R_o}\right]^2 = \frac{\hbar^2}{2mR_o^2}(n_z^2 + n_\theta^2) \tag{20}$$

In Eq. (18), the energy contribution from $\theta$-dimension is familiar. The form of the energy is $E \sim (n_\theta^2/R_o^2)$, and again, it is similar to the energy form due to the contribution of the $p^5$ in KK theory [3, 5]. This characteristic refers to the second indication that the QSPC is analogous to the KK theory.

## TIME INDEPENDENT PERTURBATION THEORY

Some quantum systems have a complicated form of a potential $\hat{V}$. This complicated form of $\hat{V}$ obscures someone who wants to figure out the wave function $\Psi_n$ and energy $E_n$ solutions. There is a method to approximate the wave function and energy solutions for any quantum system using the so-called perturbation theory [8]. The basic idea of the perturbation theory is simple. Perturbation theory tries to approximate the solution by adding some corrections to the solution of the solvable simple system [8]. In mathematical formulation, let $\hat{H}^{(0)}$ be the Hamiltonian of the solvable simple system, the total Hamiltonian $\hat{H}$ be:

$$\hat{H} = \hat{H}^{(0)} + g\hat{H}^{(1)} + g^2\hat{H}^{(2)} + \cdots \quad (21)$$

Analogous to the total Hamiltonian in Eq. (21), the wave function and the energy are in Eq. (22) and Eq. (23). The illustration of the perturbation method for solving the Schrodinger equation is in Fig. 2.

$$\Psi_n = \Psi_n^{(0)} + g\Psi_n^{(1)} + g^2\Psi_n^{(2)} + \cdots \quad (22)$$

$$E_n = E_n^{(0)} + gE_n^{(1)} + g^2E_n^{(2)} + \cdots \quad (23)$$

Here, the $g$ indicates a coupling constant, and the superscript index at $H$, $\Psi_n$, and $E_n$ indicates the order of the perturbation.

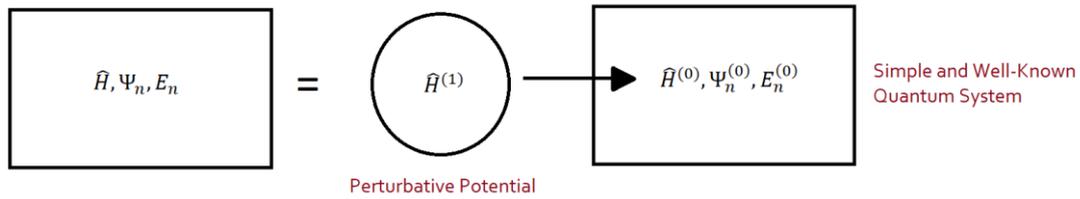

**FIGURE 2.** Illustration of the perturbation method for solving the Schrodinger equation

*First-Order Perturbation*

For the first-order perturbation, one can take the Hamiltonian in Eq. (21), the wave function in Eq. (22), and the energy in Eq. (23) as the following form:

$$\hat{H} \approx \hat{H}^{(0)} + g\hat{H}^{(1)} \quad (24)$$

$$\Psi_n \approx \Psi_n^{(0)} + g\Psi_n^{(1)} \quad (25)$$

$$E_n \approx E_n^{(0)} + gE_n^{(1)} \quad (26)$$

Eq. (24) to Eq. (26) should be the solution of the Schrodinger equation. Therefore, by inserting Eq. (24) to Eq. (26) into the Schrodinger equation, one may obtain the following relation:

$$\hat{H}\Psi_n = E_n\Psi_n \rightarrow \hat{H}^{(0)}\Psi_n^{(1)} + \hat{H}^{(1)}\Psi_n^{(0)} = E_n^{(0)}\Psi_n^{(1)} + E_n^{(1)}\Psi_n^{(0)} \quad (27)$$

By taking the scalar product of Eq. (27) with $\Psi_n^{(0)}$ and assuming $\Psi_n^{(0)}$ is an orthonormal wave function, one may obtain Eq. (28):

$$\int_{V_i} \left[\Psi_m^{*(0)}\hat{H}^{(0)}\Psi_n^{(1)} + \Psi_m^{*(0)}\hat{H}^{(1)}\Psi_n^{(0)}\right] d^3\vec{r} = \int_{V_i} \left[E_n^{(0)}\Psi_m^{*(0)}\Psi_n^{(1)} + E_n^{(1)}\Psi_m^{*(0)}\Psi_n^{(0)}\right] d^3\vec{r} \quad (28)$$

If one takes $n = m$, Eq. (28) will be reduced into Eq. (29). Eq. (29) is the so-called first-order energy correction.

$$\int_{V_i} \Psi_n^{*(0)} \widehat{H}^{(1)} \Psi_n^{(0)} d^3\vec{r} = E_n^{(1)} \qquad (29)$$

In the perturbation theory approach, the correction of the wave function should be composed of the well-known wave function (unperturbed wave function) by some linear combination constants, $c_n$, as represented in Eq. (30).

$$\Psi_n^{(1)} = \sum_n c_n \Psi_n^{(0)} \qquad (30)$$

To find $c_n$, one can use the orthonormality of the $\Psi_n^{(0)}$ as the following procedure:

$$\Psi_m^{*(0)} \Psi_n^{(1)} = \sum_n c_n \Psi_m^{*(0)} \Psi_n^{(0)} \rightarrow \int_{V_i} \Psi_m^{*(0)} \Psi_n^{(1)} d^3\vec{r} = \sum_n c_n \delta_{nm} = c_m \qquad (31)$$

$$c_m = \int_{V_i} \Psi_m^{*(0)} \Psi_n^{(1)} d^3\vec{r} \qquad (32)$$

Taking the result of Eq. (28) and the hermicity of $\Psi_n^{(0)}$, one can obtain:

$$\left(E_n^{(0)} - E_m^{(0)}\right) \int_{V_i} \Psi_m^{*(0)} \Psi_n^{(1)} d^3\vec{r} = \int_{V_i} \Psi_m^{*(0)} \widehat{H}^{(1)} \Psi_n^{(0)} d^3\vec{r} \qquad (33)$$

$$c_m = \int_{V_i} \Psi_m^{*(0)} \Psi_n^{(1)} d^3\vec{r} = \frac{1}{E_n^{(0)} - E_m^{(0)}} \int_{V_i} \Psi_m^{*(0)} \widehat{H}^{(1)} \Psi_n^{(0)} d^3\vec{r} \qquad (34)$$

Therefore, one can obtain the complete expression of the $\Psi_n^{(1)}$ in the following form (one can replace $n$ with $m$ because $n$ and $m$ here are the dummy indices):

$$\Psi_n^{(1)} = \sum_m \left[ \frac{1}{E_m^{(0)} - E_n^{(0)}} \int_{V_i} \Psi_n^{*(0)} \widehat{H}^{(1)} \Psi_m^{(0)} d^3\vec{r} \right] \Psi_m^{(0)} \qquad (35)$$

The above formulation from Eq. (28) until Eq. (35) assumes that the quantum system has a non-degenerate energy profile. For a degenerate energy profile, one can generalize $\Psi_n^{(0)}$ and $\Psi_n^{(1)}$ in the following form:

$$\Psi_n^{(0)} = c_\alpha \Psi_{n_\alpha}^{(0)} + c_\beta \Psi_{n_\beta}^{(0)} + c_\gamma \Psi_{n_\gamma}^{(0)} + \cdots \qquad (36)$$

$$\Psi_n^{(1)} = c_\alpha \Psi_{n_\alpha}^{(1)} + c_\beta \Psi_{n_\beta}^{(1)} + c_\gamma \Psi_{n_\gamma}^{(1)} + \cdots \qquad (37)$$

Here, $\alpha$, $\beta$, $\gamma$, etc., label the different states with the same energy value. Then, the coefficients $c$ ($c_\alpha, c_\beta, c_\gamma$, etc.) correspond to the linear combination coefficient of the mixture. By substituting Eq. (36) and Eq. (37) to Eq. (27) and taking the scalar product with respective labels, one can build the matrix equation as the following:

$$\begin{bmatrix} H_{\alpha\alpha} & H_{\alpha\beta} & \cdots & H_{\alpha i} \\ H_{\beta\alpha} & H_{\beta\beta} & \cdots & H_{\beta i} \\ \vdots & \vdots & \ddots & \vdots \\ H_{i\alpha} & H_{i\beta} & \cdots & H_{ii} \end{bmatrix} \begin{bmatrix} c_\alpha \\ c_\beta \\ \vdots \\ c_i \end{bmatrix} = E_n^{(1)} \begin{bmatrix} c_\alpha \\ c_\beta \\ \vdots \\ c_i \end{bmatrix} \quad (38)$$

Solving the above matrix equation (secular equation), one can obtain the first-order energy correction and coefficient $c$ ($c_\alpha, c_\beta, c_\gamma$, etc.) for degenerate cases.

## STARK EFFECT IN QUANTUM MECHANICS

One of the successful evidence of quantum mechanics as a new theoretical framework in physics during the 19th - 20th century is the ability of quantum mechanics to predict the energy level of the hydrogen atom [8]. In addition, quantum mechanics also predicts the degeneracies among those energy levels theoretically [8]. The first evidence of the degeneracies among those energy levels is from the experiment of Johannes Stark in 1913 [6]. In this experiment, Stark uses the electric field to see what happens with the energy levels of the atom or molecule. The observation shows there is energy level-splitting [6, 8]. Fig. 3 shows the Stark effect on the hydrogen and Rydberg atoms.

The energy level-splitting observed by Stark in 1913 is also well-described by the Schrodinger equation [9]. The mathematical model for the Stark effect is generally the formulation of the work needed to move a unit of electric charge (e.g., an electron, $e$) by the electric field. The Eq. (40) shows the Stark effect Hamiltonian $\widehat{H}_S$ (potential) in more detail.

$$\widehat{H}_S = \int \vec{F}_E \cdot d\vec{r} = \int (-e\vec{E}) \cdot d\vec{r} = -e\vec{E} \cdot \int d\vec{r} = -e\vec{E} \cdot \vec{r} \quad (40)$$

Here, in the Stark effect Hamiltonian, the electric field $\vec{E}$ is independent of the particle's (charge) position $\vec{r}$ but may be time-dependent. Therefore, by the presence of the Stark Hamiltonian, the total Hamiltonian for the hydrogen atom can be written in Eq. (41).

$$\widehat{H} = \left[-\frac{\hbar^2}{2m}\nabla^2 - \frac{e^2}{4\pi\epsilon_o r}\right] + \left[-e\vec{E} \cdot \vec{r}\right] \to \widehat{H}\Psi = E\Psi \quad (41)$$

The Eq. (41) can be solved by employing the perturbation theory in Eq. (38), where the Stark Hamiltonian be the perturbative potential and the system also has the degeneracies.

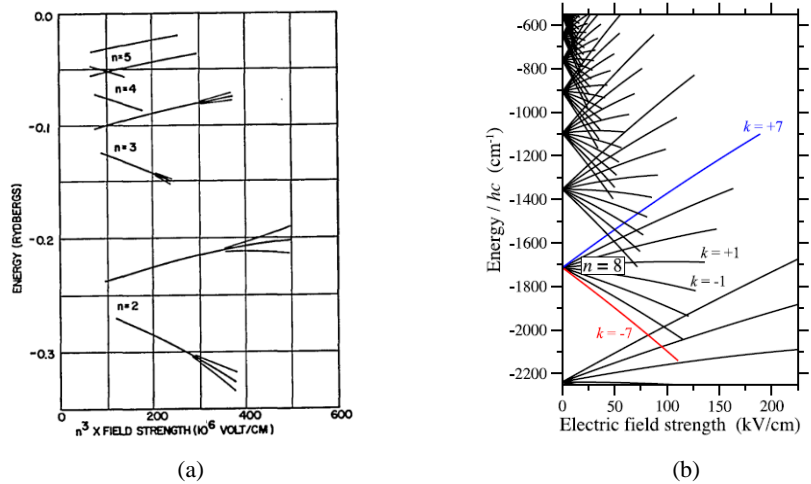

**FIGURE 3.** The observation of Stark effect in (a) the hydrogen atom [10] and (b) in the Rydberg atom [11]. Here, the energy-splitting occurs and breaks the degeneracies.

## RESULTS AND DISCUSSIONS

The main topic of this section is to see how the effect of Stark-like perturbative potential affects the energy of the QPSC up to the first-order correction. The first-order correction is considered in this calculation because this correction gives a significant contribution to the system behavior under perturbative potential.

### Stark-like Perturbative Potential and its effect on the QPSC

Stark-like perturbative potential ($\widehat{H}_{SL}$) adapts the Stark perturbative potential ($\widehat{H}_S$) in Eq. (40). However, the electron charge $e$ and the electric field $\vec{E}$ are changed [12, 13, 14]. In this paper, they are changed into any arbitrary physical constant $\beta$ and a vector function that depends on theta $\vec{V}_o(\theta) = (V_{o_x}(\theta), V_{o_y}(\theta), V_{o_z}(\theta))$. Then, the $\widehat{H}_{SL}$ for this research can be written in Eq. (42).

$$\widehat{H}_{SL} = \beta \vec{V}_o(\theta) \cdot \vec{r} \qquad (42)$$

For the sake of simplicity to see the effect of the $\widehat{H}_{SL}$ on the QPSC energy, the $\widehat{H}_{SL}$ is subjected along the $z$-axis of the system. Therefore, Eq. (42) can be rewritten into Eq. (43). In addition, for non-degenerate and degenerate cases, this research just applied the $\widehat{H}_{SL}$ to the several low-excitation spectra only. This can be seen in more detail in the description below.

$$\widehat{H}_{SL} = \beta V_{o_z}(\theta) z \qquad (43)$$

### *Non-degenerate Case*

The first-order correction of energy for the non-degenerate case $E_n^{(1)}$ is shown in Eq. (29). By taking the unperturbed wave function $\Psi_{n_z,n_\theta}^{(0)}(\theta, z, 0)$ in Eq. (15) and $\widehat{H}_{SL}$ in Eq. (43) to Eq. (29), one can determine the $E_{n_z,n_\theta}^{(1)}$ as written in Eq. (44).

$$E_{n_z,n_\theta}^{(1)} = \iint \Psi_{n_z,n_\theta}^{*(0)} \widehat{H}_{SL} \Psi_{n_z,n_\theta}^{(0)} R_o d\theta dz = \frac{\beta}{\pi R_o L} \int_0^L z \sin^2\left(\frac{n_z \pi z}{L}\right) dz \int_0^{2\pi} V_{o_z}(\theta) d\theta \qquad (44)$$

Using $\int_0^L z \sin^2\left(\frac{n_z \pi z}{L}\right) dz = \frac{1}{4} L^2$ for $n_z \in \mathbb{Z}$ and $n_z \geq 1$, then Eq. (44) can be reduced into Eq. (45) as in the following form.

$$E_{n_z,n_\theta}^{(1)} = \frac{\beta L}{4\pi R_o} \int_0^{2\pi} V_{o_z}(\theta) d\theta \qquad (45)$$

From Eq. (45), one can see that the first-order energy correction depends on the $V_{o_z}(\theta)$ only. This is the contribution from the Stark-like perturbative potential to the non-degenerate QPSC. In addition, there are no dependency on the QPSC quantum numbers $n_z$ and $n_\theta$ for the non-degenerate case. This fact implies that the perturbation from the Stark-like perturbative potential to the QPSC has a difficulty for differentiating the effect in $z$-axis or in $\theta$ direction. It can be seen easily by adding any functions of $V_{o_z}(\theta)$ to Eq. (45), the result is a constant number. The energy level shift of each pair of QPSC quantum numbers is also constant. Therefore, there are no way to interpret the Stark-like perturbative potential to the non-degenerate QPSC as a technique to uncloak the extra dimension ($x^5$) which is analogous to the cylinder circumference in the QPSC.

## *Degenerate Case*

As seen in the non-degenerate case, the Stark-like perturbative potential cannot uncloak the extra dimension ($x^5$) which is analogous to the cylinder circumference in the QPSC. In order to uncloak the extra dimension ($x^5$), let jump to the degenerate case with the same Stark-like perturbative potential. The matrix element in Eq. (38) can be determined as in Eq. (47).

$$H_{ij} = \int_{V_i} \Psi_i^{*(0)} \hat{H}_{SL} \Psi_j^{(0)} d^3\vec{r} \qquad (46)$$

$$H_{ij} = \beta R_o \int_0^{2\pi} \int_0^L \Psi_i^{*(0)} z V_{o_z}(\theta) \Psi_j^{(0)} d\theta dz \qquad (47)$$

Where $i, j = \{\alpha, \beta, \gamma, ...\}$. In the QPSC case, there are two quantum numbers ($n_z$ and $n_\theta$). By choosing $R_o = L/\pi$ as mentioned in Eq. (20), one can create the two-fold degenerate QPSC. Then, the index for $i$ and $j$ are only available for $i, j = \{\alpha, \beta\}$ and the perturbative matrix in Eq. (38) only has 2 x 2 matrix dimension.

In the degenerate QPSC, the wave function can be written as $\Psi_\alpha = \frac{1}{\sqrt{L^2}} \sin\left(\frac{n_{z_i} \pi z}{L}\right) e^{in_{\theta_i}\theta}$ and $\Psi_\beta = \frac{1}{\sqrt{L^2}} \sin\left(\frac{n_{z_j} \pi z}{L}\right) e^{in_{\theta_j}\theta}$. Here $(n_{z_j}, n_{\theta_j})$ are the permutation of $(n_{z_i}, n_{\theta_i})$ (e.g., $n_{z_i}, n_{\theta_i} = \{1,2\} \to n_{z_j}, n_{\theta_j} = \{2,1\}$). Substituting $\Psi_\alpha$ and $\Psi_\beta$ to Eq. (47), one can derive the matrix element in Eq. (38) for the degenerate QPSC as the following form. From the expression of the matrix element in Eq. (48), one can see that the matrix element value for the degenerate QPSC depends on the quantum numbers.

$$H_{ij} = \frac{\beta R_o}{L^2} \int_0^L z \sin\left(\frac{n_{z_i} \pi z}{L}\right) \sin\left(\frac{n_{z_j} \pi z}{L}\right) dz \left[ \int_0^{2\pi} V_{o_z}(\theta) e^{i(n_{\theta_j} - n_{\theta_i})\theta} d\theta \right] \qquad (48)$$

In order to see the effect of the Stark-like perturbative potential to the degenerate case of QPSC more clearly, this research taking the low-excitation energy level for the sake of simplicity. This means that this research only taking $(n_z, n_\theta) = \{(1,1), (1,2), (2,1), (2,2), (1,3), (3,1), (2,3), (3,2)\}$. By taking this set of quantum number to Eq. (48), one can summarize the matrix element for the low-excitation energy level of degenerate QPSC as in Table 1 below.

**TABLE 1.** The Perturbation Matrix Element of the degenerate QPSC under the Stark-like Perturbative Potential. Here $I_o = \int_0^{2\pi} V_{o_z}(\theta)d\theta$, $I_1 = \int_0^{2\pi} V_{o_z}(\theta)e^{-i\theta}d\theta$, and $I_2 = \int_0^{2\pi} V_{o_z}(\theta)e^{i\theta}d\theta$

| Quantum Number | Matrix Element $H_{ij}$ |
|---|---|
| (1,1) | No Degeneracy |
| (1,2), (2,1) | $H_{\alpha\alpha} = H_{\beta\beta} = \dfrac{\beta L}{4\pi}I_o$ <br> $H_{\alpha\beta} = -\dfrac{8\beta L}{9\pi^3}I_1 \; ; H_{\beta\alpha} = -\dfrac{8\beta L}{9\pi^3}I_2$ |
| (2,2) | No Degeneracy |
| (1,3), (3,1) | $H_{\alpha\alpha} = H_{\beta\beta} = H_{\alpha\beta} = H_{\beta\alpha} = 0$ |
| (2,3), (3,2) | $H_{\alpha\alpha} = H_{\beta\beta} = \dfrac{\beta L}{4\pi}I_o$ <br> $H_{\alpha\beta} = -\dfrac{24\beta L}{25\pi^3}I_1 \; ; H_{\beta\alpha} = -\dfrac{24\beta L}{25\pi^3}I_2$ |

By employing the result in Table 1 and solving the secular equation in Eq. (38), one can derive the first-order energy correction for the degenerate QPSC under the Stark-like perturbative potential as shown in Eq. (51). The Table 2 is also summarized the First-order energy correction for the degenerate QPSC under the Stark-like perturbative potential for $(n_z, n_\theta) = \{(1,1), (1,2), (2,1), (2,2), (1,3), (3,1), (2,3), (3,2)\}$. Fig. 4 illustrates the energy level splitting of the degenerate QPSC under the Stark-like perturbative potential.

$$\begin{bmatrix} H_{\alpha\alpha} & H_{\alpha\beta} \\ H_{\beta\alpha} & H_{\beta\beta} \end{bmatrix} \begin{bmatrix} c_\alpha \\ c_\beta \end{bmatrix} = E^{(1)} \begin{bmatrix} c_\alpha \\ c_\beta \end{bmatrix} \tag{49}$$

$$\begin{bmatrix} H_{\alpha\alpha} - E^{(1)} & H_{\alpha\beta} \\ H_{\beta\alpha} & H_{\beta\beta} - E^{(1)} \end{bmatrix} \begin{bmatrix} c_\alpha \\ c_\beta \end{bmatrix} = \begin{bmatrix} 0 \\ 0 \end{bmatrix} \tag{50}$$

$$E^{(1)}_{\alpha,\beta} = \frac{1}{2}(H_{\alpha\alpha} + H_{\beta\beta}) \pm \frac{1}{2}\sqrt{(H_{\alpha\alpha} + H_{\beta\beta})^2 - 4(H_{\alpha\alpha}H_{\beta\beta} - H_{\beta\alpha}H_{\alpha\beta})} \tag{51}$$

**TABLE 2.** First-order Energy Correction of the degenerate QPSC under the Stark-like perturbative potential

| Quantum Number | First-order Energy Correction |
|---|---|
| (1,1) | No Degeneracy |
| (1,2), (2,1) | $E^{(1)}_{\alpha,\beta} = \dfrac{\beta L}{4\pi}I_o \pm \dfrac{8\beta L}{9\pi^3}\sqrt{I_1 I_2}$ |
| (2,2) | No Degeneracy |
| (1,3), (3,1) | $E^{(1)}_{\alpha,\beta} = 0$ |
| (2,3), (3,2) | $E^{(1)}_{\alpha,\beta} = \dfrac{\beta L}{4\pi}I_o \pm \dfrac{24\beta L}{25\pi^3}\sqrt{I_1 I_2}$ |

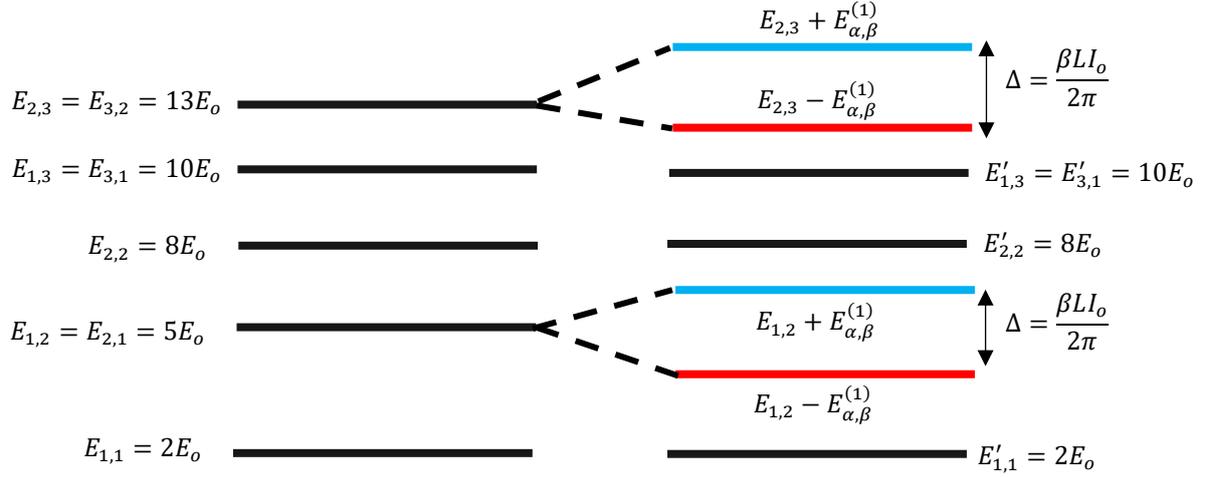

**FIGURE 4.** The illustration of the degenerate QPSC energy level shift due to the Stark-like Perturbative Potential.

From the result of Table 1, Table 2, and Fig. 4 are the best indication that the degenerate QPSC under the Stark-like perturbative potential may open the opportunity to uncloak the extra dimension $x^5$ in KK theory (which is analogous to the $\theta$ variable in the QPSC). It is because the quantum numbers $(n_z, n_\theta)$ are significantly determining the existence of the perturbation terms (that are depicted by $H_{\alpha\beta}$ and $H_{\beta\alpha}$) which correspond to the splitting of the energy level. The splitting of the energy level here is interpreted as the uncloaked extra dimension. The other factor that determines the existence of the perturbation terms is the expression of $V_{o_z}(\theta)$. If the strength of the $V_{o_z}(\theta)$ is large enough, the spliting in the energy level is also becoming larger. In addition, there are two remarks related to the Stark-like perturbative potential application to the QPSC.

*Remark 1: Perturbative Splitting Rule*

Let back to the Eq. (48). The first term of Eq. (48) shows the integral $\int_0^L z \sin\left(\frac{n_{z_i}\pi z}{L}\right) \sin\left(\frac{n_{z_j}\pi z}{L}\right) dz$. This integral is purely depends on the position in $z$-axis and the quantum number in $z$-direction. This kind of integral is analogous to the dipole interaction with the dipole moment $M$ [15, 16, 17, 18] as shown in Eq. (52).

$$M \propto \int z \psi_i^*(z) \psi_j(z) \, dz \tag{52}$$

where the index $i$ and $j$ are the corresponding quantum numbers of the state $\psi$. In the case of dipole moment, the selection rule of the dipole transition [17, 18] is more important to figure out and can be obtained easily by checking which quantum numbers that make Eq. (52) has non zero value ($M \neq 0$). In the Stark-like perturbative potential to the QPSC, the energy level splitting exist if and only if the difference of the $n_{z_i}$ and $n_{z_j}$ is $\pm 1$. This leads to the so-called "perturbative splitting rule" to the QPSC and should be considered also in the effort to uncloak the extra-dimension.

$$\Delta n_z = \pm 1 \tag{53}$$

*Remark 2: Strategy for choosing the $V_{o_z}(\theta)$*

The existence of Eq. (53) as the splitting rule forces one who has to uncloak the extra-dimension using the Stark-like perturbative potential to choose the appropriate $V_{o_z}(\theta)$ which satisfies Eq. (54). Eq. (54) ensures the splitting can be observed by any experimental measurements.

$$\{I_1, I_2\} \in \mathbb{R} \wedge \{I_1, I_2\} \neq 0 \tag{54}$$

The only possible $V_{o_z}(\theta)$ that satisfies Eq. (54) is a periodic potential in $\theta$-direction like shown in Eq. (55). The $V_o$ and $\gamma$ are the non zero constants (with $\gamma = \{\frac{1}{2}, \frac{3}{2}, \frac{5}{2}, \frac{7}{2}, ...\}$). This statement can be proven by taking the general form of a polynomial function in $\theta$-variable to the $I_1$ and $I_2$.

$$V_{o_z}(\theta) = V_o \cos \theta \vee V_{o_z}(\theta) = V_o \sin \gamma\theta \tag{55}$$

<u>Proof</u>: Let consider $V_{o_z}(\theta) = \sum_{k=0}^{\infty} a_k \theta^k$, then by the definition of $I_1$ and $I_2$, one can have these following results (Eq. (56) and Eq. (57)). The result shows that the integral values are a complex number.

$$I_1 = \int_0^{2\pi} \left[\sum_{k=0}^{\infty} a_k \theta^k\right] e^{-i\theta} d\theta = \sum_{k=0}^{\infty} a_k \left[\int_0^{2\pi} \theta^k \cos\theta \, d\theta - i \int_0^{2\pi} \theta^k \sin\theta \, d\theta\right] \in \mathbb{C} \tag{56}$$

$$I_2 = \int_0^{2\pi} \left[\sum_{k=0}^{\infty} a_k \theta^k\right] e^{i\theta} d\theta = \sum_{k=0}^{\infty} a_k \left[\int_0^{2\pi} \theta^k \cos\theta \, d\theta + i \int_0^{2\pi} \theta^k \sin\theta \, d\theta\right] \in \mathbb{C} \tag{57}$$

Then, for the periodic potential $V_{o_z}(\theta) = V_o \cos\theta$,

$$I_1 = \int_0^{2\pi} [V_o \cos(\theta)] e^{-i\theta} d\theta = V_o \pi \in \mathbb{R} \tag{58}$$

$$I_2 = \int_0^{2\pi} [V_o \cos(\theta)] e^{i\theta} d\theta = V_o \pi \in \mathbb{R} \tag{59}$$

and for the $V_{o_z}(\theta) = V_o \sin \gamma\theta$,

$$I_1 = \int_0^{2\pi} [V_o \sin(\gamma\theta)] e^{-i\theta} d\theta = \frac{2V_o \sin(\gamma\pi)}{\gamma^2 - 1} (\gamma \sin(\gamma\pi) - i\cos(\gamma\pi)) \in \mathbb{R} \text{ if } \gamma \in \left\{\frac{1}{2}, \frac{3}{2}, \frac{5}{2}, ...\right\} \tag{60}$$

$$I_2 = \int_0^{2\pi} [V_o \sin(\gamma\theta)] e^{i\theta} d\theta = \frac{2V_o \sin(\gamma\pi)}{\gamma^2 - 1} (\gamma \sin(\gamma\pi) + i\cos(\gamma\pi)) \in \mathbb{R} \text{ if } \gamma \in \left\{\frac{1}{2}, \frac{3}{2}, \frac{5}{2}, ...\right\} \tag{61}$$

These kind of periodic potentials in Eq. (55) as an implication of the pertrubative splitting rule may also give an alternative insight regarding the physics that uses the extra dimension (KK theory for example). These physics include the UV completion in multi-natural inflation model and SUSY breaking model [19], and also the implementation of optical lattice technology to simulate $D+1$ - dimensional quantum system using $D$-dimensional [20].

## CONCLUSION

The Kaluza-Klein (KK) Theory is one of the theories which capables to unify the gravitational interaction and the electromagnetic interaction by introducing the extra dimension ($x^5$). The extra dimension ($x^5$) in the KK theory has a small size. Therefore, this extra dimension is somewhat cloaked and becomes the best argument to understand why the gravitational interaction is weaker than the electromagnetic interaction. This research has explored the way how to uncloak the extra dimension in the KK theory taking the perturbative potential as the Stark-like perturbative potential $\hat{H}_{SL} = \beta z V_{o_z}(\theta)$ to the non-relativistic quantum particle on the surface of the cylinder (with radius $R_o$ and length $L$) QPSC. The QPSC is taken because the $\theta$-variable in this system acts like the extra dimension in KK theory. Suppose that the QPSC is in the non-degenerate case, the Stark-like perturbative potential only shift the energy level of the QPSC and cannot differentiate which one is the extra dimension. Meanwhile, the if the QPSC is in the degenerate case ($R_o = L/\pi$), the Stark-like perturbative potential can split the energy level and the $\theta$-variable contribution can be differentiate easily. These findings may given the insight for how to uncloak the extra dimension in KK theory which corresponds to the gravitational interaction and can be validated by the approriate experimental technique. For the future works, the comparison of the Stark-like perturbative potential to the other types of perturbative potential like Zeeman-like, Anharmonic-like, and so on can be proceed. In order to see, which the most effective well-known perturbative scheme that may uncloak the extra dimension effect.